# Personalized Survival Predictions for Cardiac Transplantation via Trees of Predictors

**Authors:** J. Yoon[1], W. R. Zame[1,2,3], A. Banerjee[3], M. Cadeiras[1], A. M. Alaa[1], M. van der Schaar[1,4,5,*]

**Affiliations:**

[1]University of California, Los Angeles, CA, USA.

[2]Nuffield College, Oxford University, Oxford, UK.

[3] Farr Institute of Health Informatics Research, University College, London, UK

[4]Oxford Man Institute, Oxford University, Oxford, UK.

[5]Alan Turing Institute, London, UK.

*Corresponding author (E-mail): mihaela.vanderschaar@eng.ox.ac.uk

**One Sentence Summary:** This paper develops a novel machine learning methodology (ToPs – Trees of Predictors) that discovers clusters of patients and creates cluster-specific predictive models of survival for patients in those clusters, and then applies the methodology to the United Network for Organ Sharing data set to demonstrate significantly improved personalized predictions of survival in patients with advanced heart failure before and after heart transplantation.

**Abstract**: Given the limited pool of donor organs, accurate predictions of survival on the wait list and post-transplantation are crucial for cardiac transplantation decisions and policy.  However, current clinical risk scores do not yield accurate predictions. We develop a new methodology (ToPs – Trees of Predictors) built on the principle that specific predictors should be used for specific clusters within the target population.  ToPs *discovers* these specific clusters of patients and the specific predictor that perform best for each cluster. In comparison with current clinical risk scoring systems, our method provides significant improvements in the prediction of survival time on the wait list and post-transplantation.  For example: in terms of 3-month survival for patients who were on the US patient wait-list in the period 1985-2015, our method achieves AUC of 0.847; the best commonly used clinical risk score (MAGGIC) achieves 0.630. In terms of 3-month survival/mortality predictions (in comparison to MAGGIC), holding specificity at 80.0%, our algorithm correctly predicts survival for 1,228 (26.0%) more patients out of 4,723 who actually survived; holding sensitivity at 80.0%, our algorithm correctly predicts mortality for 839 (33.0%) more patients out of 2,542 who did not survive. Our method achieves similar improvements for other time horizons and for predictions post-transplantation.  Therefore, we offer a more accurate, personalized approach to survival analysis that can benefit patients, clinicians and policymakers in making clinical decisions and setting clinical policy. Because risk prediction is widely used in diagnostic and prognostic clinical decision-making across diseases and clinical specialties, the implications of our methods are far-reaching.

# INTRODUCTION

Heart transplantation is the treatment of last resort for patients suffering from end-stage heart failure (1-3). For such patients, a transplant represents the opportunity for improved life expectancy and quality of life. However, the demand for hearts (the number of patients on the wait-list) exceeds the supply (the number of donors) so that patients (potential recipients) are at substantial risk of dying while on the waiting list. In the U.S. between 1985 and 2015, more than 36,000 patients were on a waiting list for a heart transplant without obtaining one; of these, more than 15,000 (41.6%) died within a year of entering the wait list (4-6). In the same period, more than 60,000 patients received a heart transplant; of these, more than 8,000 (13.3%) patients died within one year.

Minimizing mortality on the wait list and maximizing survival after transplant are difficult problems (7-15) but are crucial for effective transplantation practice and policy (16, 17). Dealing effectively with these problems requires accurate predictions of survival for patients on the wait list and for patients who receive transplants. However, the clinical risk scores that are currently used to predict survival on the wait list and survival after transplantation perform rather poorly: for survival on the wait list, none of them achieves AUC (Area Under the receiver operating characteristic Curve) of 0.650 at time horizons of 3 months, 1 year, 3 years or 10 years; for survival post-transplantation, none of them achieves AUC of 0.600 for the same time horizons. This poor performance results (at least in part) from the inability of the current clinical methods to properly address three complex aspects:
- The populations of patients and donors are extremely heterogeneous and the effect of particular features (covariates) on survival is different for different subpopulations.
- The dependence of survival on features involves the interactions *between* features – in particular, dependence is non-linear – and the interactions are different for different subpopulations.
- The features and interactions between features that have the most effect on survival depend on the time horizon: the features and interactions between features that are most important for survival for 1 year are different from those that are most important for survival for 3 months or 3 years or 10 years.

Current clinical risk scores fail to address any of these complex aspects. They ignore the heterogeneity of the populations and employ a one-size-fits-all methodology. They ignore potential interactions between features and compute a risk score that is linear in scores associated to individual features. They ignore the differences between time horizons, computing only a single risk score.

Our work develops and applies novel machine learning methods for survival analysis to craft a methodology, *Trees of Predictors* (*ToPs*), and a particular implementation, *ToPs/Regression (ToPs/R),* that is expressly designed to address all these complex aspects. We address the heterogeneity of the patient and donor populations by endogenously identifying specific subpopulations (*clusters*) and applying different predictors to different subpopulations. We address the interactions between features by constructing non-linear predictors for some subpopulations and address the different importance of features for different subpopulations by choosing different coefficients in the predictors constructed for those subpopulations. We address the importance of different features and different interactions between features at different time horizons by tuning the construction to the particular time horizon. Our method is interpretable and applies to survival both on the wait list and post-transplantation – and so provides an end-to-end methodology for use by clinicians. Our underlying construction may be carried out entirely offline, allowing the calculations for a particular patient to be carried out in real-time. Our website provides a convenient clinical decision support system that is readily usable in clinical settings: http://medianetlab.ee.ucla.edu/ToPs_TransplantSurvival. Our method yields significantly more accurate predictions than current clinical methods (and more accurate predictions than alternative machine learning methods that are not tailored for the particular application) for both survival on the wait-list and survival post-transplantation, and for all considered time horizons.

# RESULTS

**Performance improvement of ToPs/R**

Tables 1 and 2 show the prediction performance in wait-list and post-transplantation for our method (ToPs/R), current clinical risk scores, familiar regression models and several machine learning benchmarks. (We conducted our study using the United Network for Organ Sharing database of patients who were registered to undergo heart transplantation during the years from 1985 to 2015; see Methods and Methodology for a more complete description of the data.) As can be seen, our method provides very large improvements in the prediction of survival time on the wait list and post-transplantation in comparison with the clinical risk scores. Furthermore, our method also shows consistently and statistically significantly (p-value < 0.01) better performance than the machine learning benchmarks. For instance, for wait-list survival at a time horizon of 3 months, our method achieves Area Under the receiver operating characteristic Curve (AUC) of 0.847 (95% Confidence Interval (CI): 0.846-0.848). By contrast, the best performing currently used clinical predictor achieves AUC of only 0.630 (CI: 0.629-0.632). The improvements achieved by our algorithm are similar both for other time horizons and for post-transplantation predictions.

Absolute increase (or percentage increase) in AUC is a measure of improved performance, but understates the performance improvement. A predictor performs well if the region under the Receiver Operating Characteristic (ROC) curve is large; or equivalently if the area *above* the ROC curve is small. AUC measures the area of the region under the ROC curve, but a more revealing measure of performance improvement is in terms of *predictive loss* – the area of the region *above* the ROC curve; this is just the difference 1-AUC. We therefore express performance gain of our method over a competing method as the *reduction in the predictive loss* expressed as a percentage. As Table 3 demonstrates, for survival in the wait list ToPs/R out-performs the best clinical risk scores by 35.2-58.6%, the basic regression methods (on which ToPs/R is built) by 18.6-36.3%, and the best machine learning benchmarks by 5.5-12.8% at all time horizons; for survival post-transplantation, the improvements are smaller but still very substantial. All the improvements are statistically significant (p-value < 0.01).

Figures 1(a) and 1(b) provide visual evidence of the extent of the improvement of our method over competing methods. Figure 1(a) shows the actual ROC curves for ToPs/R, for the best machine learning algorithm (Random Forest), and for the best clinical risk score (Meta-Analysis Global Group in Chronic Heart Failure (MAGGIC)) for 3-month survival on the wait list; Figure 1(b) shows the actual ROC curves for 3-month survival post-transplantation (18).

An alternative illustration of the improvement achieved by our method is the increase in the *number* of correct predictions. For 3-month survival in the wait list, holding specificity at 80.0%, ToPs/R correctly predicted survival for 1,228 *more* patients (26.0% of the 4,723 patients survived up to 3 months) and, holding sensitivity at 80%, ToPs/R correctly predicted mortality for 839 *more* patients (33.0% of the 2,542 patients who died within 3 months), in comparison with the best clinical risk score, MAGGIC. For 3 month survival post-transplantation, with the same specificity and sensitivity,

ToPs/R correctly predicted survival for 1,401 *more* patients (13.0% of the 10,777 patients who survived up to 3 months), and correctly predicted mortality for 158 more patients (14.0% of 1,129 patients who died within 3 months), in comparison with the best clinical risk score (the Index for Mortality Prediction After Cardiac Transplantation (IMPACT)) (19). Because survival (or mortality) in the wait list (urgency) and survival post-transplantation (benefit) are perhaps the most important factors in transplant policy, these predictive improvements are of great importance.

**Predictive features**

Some features are more predictive of survival than other features, and some features are more predictive for survival for a particular time horizon than for other time horizons. Figure 2 presents heat maps displaying the predictive value of the features for various predictions over different time horizons. For instance, as can be seen in Figure 2(f), donor's age is a very predictive feature for long-term post-transplantation survival, but it is less predictive of short-term survival; the most predictive features for short-term survival are the need for advanced cardiac and respiratory life support (ECMO and ventilator support, etc.). Moreover, the predictive power of donor's age also differs across various subpopulations as well as across different time horizons. For instance, as Figure 3 shows, ToPs/R splits the entire population according to whether the creatinine level is below/above 4.64 mg/dL. The set of patients for whom the creatinine level is above 4.64 mg/dL is deemed Cluster 17; ToPs/R does not split this cluster any further. To see the importance of features in Cluster 17, refer again to the heat map Figure 2(d): we see that comorbidities such as diabetes are much more predictive of long-term survival than the donor's age. This is consistent with the fact that chronic kidney disease and its interrelated set of comorbidities (which includes diabetes) can generally worsen cardiovascular outcomes (20-21). Diabetes does not have the same predictive power for other groups of patients. For example, in Cluster 5 (patients with creatinine level below 1.74 mg/dL, Panel Reactive Antibody (PRA) above 27% and BMI within the "normal" range 23.0-27.3 kg/m, the donor's age and the ischemic time have more predictive power than diabetes. In this particular example, creatinine serves as a discriminative feature that filters out populations for whom survival prediction needs to consider different predictive features via "customized" predictive models. This example also sheds light on how ToPs/R can recognize the impact of comorbidities – in this case renal failure and diabetes – on a patient's survival: ToPs/R recognizes some features related to comorbidities as being discriminative, and also learns the appropriate predictive features for patients with those comorbidities. (Note that two of the clinical risk scores – Donor Risk Index (DRI) and IMPACT – do not use diabetes as an input to their prediction rule; this may lead to misinformed surgical decisions for patients with comorbidities (19, 22).)

**LVADs**

Figure 4(a) shows the survival curves (using Kaplan-Meier estimators) for patients who received an LVAD (left ventricular assist device) in the periods 2005-2009 and 2010-2015, the survival curves for patients who had similar features (determined by Propensity Score Matching (23)) – referred to as "LVAD-eligible" – but who did not receive an LVAD in the same periods, and the survival curves for the overall population in the same periods. For patients who received an LVAD in the period 2005-2009, survival probabilities at 3 months, 1 year and 3 years are approximately 0.896, 0.777, 0.626 (respectively); for patients who received an LVAD in the period 2010-2015, survival probabilities at the same time horizons are approximately 0.931, 0.831, 0.640 (respectively). (Because most patients who received an LVAD did so after 2005, there is not enough data for meaningful analysis of 10-year survival.) In comparison with LVAD-eligible patients, those who actually received an LVAD experienced much superior survival probabilities at all time horizons. In terms of *predictions* of survival of LVAD patients on the wait-list, ToPs/R yields AUC of 0.826, 0.821, 0.796 at time horizons of 3-months, 1-year, 3-years. This performance is similar to the performance of ToPs/R for the population as a whole and superior to all the clinical risk scores and machine learning benchmarks.

Figure 4(b) shows the survival curves for post-transplantation patients who received an LVAD in the periods 2005-2009 and 2010-2015, the survival curves for post-transplantation "LVAD-eligible" patients in the same periods, and the survival curves for the overall population of post-transplantation patients in the same periods. As can be seen, survival post-transplantation is not very different for the various groups of patients. But when we compare Figures 4(a) and 4(b), it can also be seen that survival is much better for patients who received a heart transplant (survival probabilities at 3 months, 1 year and 3 years are approximately 0.926, 0.881, 0.808, respectively) than for patients who were maintained on LVAD (survival probabilities at the same time horizons are approximately 0.931, 0.831, 0.640, respectively).

# DISCUSSION

In this study, we develop a methodology for personalized prediction of survival for patients with advanced heart failure while on the wait-list and after heart transplantation. Our methods and associated findings are important because they outperform the clinical risk scores currently in use and also because they automate the discovery and application of cluster-specific predictive models and tune predictions to the specific patient for which the prediction is to be made. The clinical and public health implications of our findings are broad and include improved personalization of clinical assessments, optimization of decision making to allocate limited life-saving resources and potential for healthcare cost reduction across a range of clinical problems.

**Key Findings**
We emphasize three key findings of our study:
- Our method significantly outperforms existing clinical risk scores, familiar regressions and state-of-the-art machine learning benchmarks in terms of accurate prediction of survival on the wait-list and post-transplantation.
- This improvement in performance has clinical significance: our method correctly predicts both survival and mortality for a larger number of patients.
- There is substantial heterogeneity – both across clusters of patients and across different time horizons. Our method captures this heterogeneity far better than other methods.

**Performance improvement of ToPs/R**
As the results show, ToPs/R achieves large performance improvements over current clinical risk-scoring models. It does so by explicitly addressing the weaknesses of these clinical models:
- Tops/R addresses the *heterogeneity of the population(s)* by identifying subpopulations (clusters) and the specific predictors that are best suited to prediction in each subpopulation. Tops/R makes predictions that are *personalized* to the features of the patient (or patient and donor).
- Tops/R addresses the *interactions between features* by using non-linear predictors for those subpopulations (clusters) in which the interactions between features are important.
- Tops/R addresses the *heterogeneity across time horizons* by constructing different predictions for different time horizons.

**Predictors and predictive clusters**
The distinction between *discriminative* and *predictive* features is useful both for understanding the working of ToPs/R and also for understanding the complex interactions between the different predictive features and their impact on survival probabilities. Discriminative features break down the entire population into smaller subpopulations (clusters). Because we use Cox Regression, Linear Regression and Logistic Regression as the basic learners, the most predictive features in a cluster are those whose coefficients have the largest magnitude (either positive or negative). Figure 5 (which shows the entire Tree of Predictors for 3-year survival in the wait-list) presents a good illustration of the point that different features are more predictive in different clusters. In each cluster, we show the (index of the) three features that are most predictive of survival (for that time horizon) for that cluster ordered in terms of predictive importance. Note that different features are important in different clusters. See also Figure 2(b) and the discussions in the Results section and in the Supplementary Materials.

The key idea of ToPs is to *discover* how to apply *different* predictors (predictive models) on *different* portions of the feature space (clusters) and hence to discover how to tune the overall prediction to the features of the patient for which the prediction is to be made. Each of the predictors applied is constructed from one of a prescribed set of base learners (algorithms). In principle, the prescribed set of base learners could be arbitrary, but for the purposes of this paper we restrict to a small family of familiar base learners: Cox Regression, Linear Regression and Logistic Regression; in addition to familiarity, this has the advantage of interpretability. (Expanding the set of base learners would increase the accuracy of our predictions – but at the cost of interpretability; this represents a familiar trade-off.) To construct a predictor from a base learner means to choose the parameters of the learner; for these learners this means choosing the coefficients of the regressor. At each step of the recursive procedure, we jointly choose the base learner and the parameters of the learner in order to achieve the best fit on an appropriate portion of the data that we use for training. (See Materials and Methods for a more complete description.)

It is important to understand what it means to apply different predictors on different clusters and why doing so yields improved predictions and is key for personalized medicine. Two examples will illustrate this.

- Suppose first that we have two distinct clusters $C_1$ and $C_2$ and that we apply Linear Regression as the base learner on both of these clusters. In general, we will choose the coefficients of the linear regressor – the predictor – to best fit *different* portions of the data; hence, there will be at least one feature whose coefficient in the predictor for $C_1$ is different from its coefficient in the predictor for $C_2$. This means that this feature will have *different importance* in the predictions for patients whose features lie in $C_1$ than for patients whose features lie in $C_2$.

- Suppose instead that that we again have two distinct clusters $C_1$ and $C_2$ but that we apply Linear Regression as the base learner on $C_1$ but apply Logistic Regression as the base learner on $C_2$. By definition, Linear Regression produces a *linear* predictor, which implies that the various features *do not interact* with each other. To the contrary, Logistic Regression produces a *non-linear* predictor, which implies that the various features *do interact* with each other. (We can see similar contrasts with Cox Regression, which also produces a non-linear predictor in which the various features interact with each other – but in a way that is different from Logistic Regression.)

In short: using different predictors on different clusters means that we allow different features to have different importance and/or to interact differently for different clusters of patients. And because we make the choice of predictors endogenously and optimally, *we are letting the data tell us which choices to make*.

**Decision trees**
Although our method constructs trees of subsets of the feature space, it is very important to distinguish our method from familiar decision trees. Both our method and decision trees use a feature to create each split and hence in both cases successive splits lead to clusters in which the features are more homogeneous than in the population as a whole. However, this is really the only similarity. Decision trees split according to a feature, but they *decide* which split should be chosen according to which split leads to greater homogeneity in the *labels* – the *outcomes*. Our method splits according to a feature, but it *decides* which split should be chosen according to which split leads to more accurate *predictions*. In particular, a decision tree assigns a *single prediction* to each terminal node; our method assigns a *single predictor* to each decision node – but that single predictor may make a wide variety of predictions for the wide variety of patient features represented in that terminal node. See the Supplementary Materials for illustrative histograms.

**Survival analysis**
The existing clinical risk scores do not produce individual survival curves. Clinicians using these scores infer survival curves by clustering patients with similar scores and constructing Kaplan-Meier curves from the actual survival times for these clusters. Our method produces individual survival curves by interpolating the predictions for 3 months, 1 year, 3 years and 10 years. Our approach can be viewed as providing a non-parametric survival model. In terms of hazard functions, our method could be interpreted as forming clusters, assigning a model to each cluster, and then aggregating those models to construct a non-proportional hazard function. In contrast to familiar approaches (24, 25) our approach *learns* which clusters to create, which models to assign and how to aggregate these models.

Cox Regression produces individual survival curves. However, because Cox Regression assumes that relative hazard rates are constant over time, the survival curves produced by Cox Regression for two different patients cannot cross: if the survival probability for patient 1 is greater than that of patient 2 at a time horizon of 3 months it will also be greater at every time horizon. However, actual survival curves *may cross*. Our method allows for this is a *virtue*, reflecting the fact that the features and interactions that are most important for survival at 3 months are different from the features and interactions that are most important for survival at longer horizons. (See Figure 2.)

**LVADs**
Perhaps the most significant advance in heart failure technology has been the development of mechanical circulatory support devices, especially LVADs that can be surgically implanted and support or replace the function of the failing heart. Patients who receive LVADs typically remain on LVADs until transplantation or death; a small proportion of patients recover their heart function sufficiently that their LVAD can be explanted. First generation (pulsatile) LVADs were approved by the U.S. Food and Drug Administration in 1994. However, use of LVADs only increased significantly in about 2005 and became widespread in the U.S. and some other countries in 2009 with the development of second generation (non-pulsatile) LVADs (26-28). Initially, LVADs were used largely as a bridging therapy, maintaining patients until they could receive a transplant. After the landmark REMATCH trial in 2001 (29), it has become more common to use LVADs as destination therapy for patients with end-stage heart failure who are not eligible for Heart Transplantation (30, 31). Approximately 25.0% of patients currently on wait lists in the U.S. now receive LVADs. Therefore, there is a need to assess whether different risk prediction tools are required for this patient subpopulation. In this paper, we analyze the subgroup of patients who receive an LVAD, both with respect to survival in the wait-list and post-transplantation. Our method achieves similar predictive accuracy for patients who received LVADs (both in the wait-list and post-transplantation) and patients who did not; these predictions are again significantly more accurate than existing clinical scores and other machine learning methods. Because our methods are applicable both to patients who have received LVADs and those who have not, and because our methods predict *both* survival on the wait-list and survival post-transplantation with significantly improved accuracy, our methods provide the potential for improved heart transplantation policy that could lead to improved clinical outcomes.

**Clinical support**

Our work provides support for clinical decision making *in real time*. By using our user-friendly website http://medianetlab.ee.ucla.edu/ToPs_TransplantSurvival and entering relevant features, the clinician can obtain immediate predictions for a specific patient, including survival on the wait-list, impact of an LVAD (if relevant), and benefit of transplantation (if relevant). All of this can be done at the desk of the clinician or the bedside of the patient, in no more time than is currently required to access the clinical risk scores, and with much greater accuracy (and confidence).

**Risk scoring and usability**

Our analysis shows that our method provides significantly greater predictive power for survival while on the wait-list and post-transplantation. However, more research in actual practice is required. This success of our method in the setting of cardiac transplantation suggests it may have wide applicability and usability for risk prognosis and diagnosis for other medical conditions and diseases. Such wider applications may benefit from further refinements of our method.

## MATERIALS AND METHODS

**Study Design**
We conducted our study using the United Network for Organ Sharing (UNOS) database of patients who were registered to undergo heart transplantation during the years from 1985 to 2015 (available at https://www.unos.org/data/). This provided a dataset of 60,400 patients who received heart transplants and 36,329 patients who were on a wait list but did not receive heart transplants; we refer to the former as transplanted patients and to the latter as wait-listed patients. Of the 60,400 transplanted patients, 29,436 patients (48.7%) were followed until death and the remaining 30,964 patients (51.3%) were either still alive or their exact survival time was not known (i.e., these patients were right-censored). Of the 36,329 wait-listed patients, 18,584 patients (51.2%) were followed until death; the remaining 17,745 patients (48.8%) were either still alive or their exact survival time was not known (i.e., these patients were right-censored).

Patients in the dataset are described by a total of 504 clinical features. Of these, 334 features pertain to (potential) recipients, 150 features pertain to donors and 20 features pertain to donor-recipient compatibility. We discarded 12 features that can be obtained only after transplantation and so cannot be used for prediction before transplantation, and 439 features for which more than 10.0% of the information was missing. After discarding these, we were left with 33 recipient features, 14 donor features and 6 donor-recipient compatibility features – a total of 53 features (See Supplementary Materials.) Categorical binary features (e.g. Male/Female) are represented as 0, 1; categorical non-binary features are converted to binary features (e.g. blood type A/not-A); other features are represented as real numbers. For patients and features for which values are missing, we use standard imputation methods. More specifically, we conduct 10 multiple imputations using Multiple Imputation by Chained Equations (MICE) as in (32) and represent the performance of the various algorithms by showing both means and confidence intervals. Furthermore, we use the Sequential Correlated Feature Selection (SCFS) procedure of (33) to compute a relevance score for each feature. These scores depend on the task at hand: the relevance scores for survival on the waiting list are different for different time horizons, and the relevance scores for survival on the waiting list are different from the relevance scores for survival after transplantation (for various time horizons). The Supplementary Materials provides a complete list of all 53 features, the relevance scores for each prediction we make, and the subset of features actually used for that prediction.

To develop and test our models we used 5-fold cross validation. We randomly separated both the wait list and post-transplantation data sets into 5 folds. In each of 5 iterations, 4 of the folds (80.0% of the data) were used for development of the models and the remaining fold (20.0% of the data) was put aside for evaluating performance.

**Clinical risk scores and survival analysis**
Three predictors for survival times on the wait list are in widespread use in clinical practice: the Heart Failure Survival Score (HFSS), the Seattle Heart Failure Model (SHFM) and Meta-Analysis Global Group in Chronic Heart Failure (MAGGIC) (18, 34, 35). Similarly, three predictors for survival times after transplant are in widespread use in clinical practice: the Donor Risk Index (DRI), the Risk-Stratification Score (RSS), and the Index for Mortality Prediction After Cardiac Transplantation (IMPACT) (19, 22, 36). All of these predictors compute an overall risk score (a weighted sum of scores of individual features). The numerical score assigned to a particular patient is only meaningful in comparison with the numerical scores assigned to other patients: patients with higher risk scores are viewed as more likely to die/less likely to survive for any given time horizon.

**Performance metrics and comparisons**
The performance of any risk scores can be assessed in terms of the *Receiver Operating Characteristic* (ROC) curve that plots the *True Positive Rate* (TPR) (*sensitivity*) against the False Positive Rate (FPR) (1 – *specificity*). A summary performance statistic is the *Area Under the ROC Curve* (AUC). We report the performance (AUC for survival on the wait list and for survival post-transplantation at the time horizons of 3 months, 1 year, 3 years and 10 years) of ToPs/R, of three familiar regression methods, and of the relevant clinical risk scores. We also compare with a number of state-of-the-art machine learning approaches.

**Implementation of ToPs**
Our approach uses a novel ensemble learning method that we call *ToPs* because it constructs *Trees of Predictors*. The implementation we use here uses as its base learners (algorithms) three familiar regression algorithms – Cox Regression, Linear Regression and Logistic Regression – so we use the acronym *ToPs/R*. In fact, we could use *any* collection of base learners, and a larger or more sophisticated collection of base learners would lead to improved predictive accuracy. The choice of base learners made here was motivated by their familiarity and widespread use and because it lends itself more readily to interpretation of the clusters that are discovered. This is just another example of the well-known trade-off between accuracy and interpretability.

**Model development: Base learners**
As discussed in the Introduction, the current clinical models compute risk scores as weighted linear combinations of scores of various features. However, as Tables 1 and 2 show, the current clinical models do not perform well. Because the various clinical models do not *optimize* the choices of features and weights used and compute a single risk score to be used for all time horizons, there is room for significant improvement by carrying out a Linear Regression that fits the *best* linear regressor for each time horizon separately. As Tables 1 and 2 show, this approach leads to a substantial improvement over all of the clinical risk scores. However, by its very nature, Linear Regression does not capture interactions *between* features. Both Cox Regression and Logistic Regression, *do* capture (some) interactions between features. As Tables 1 and 2 show, however, although Logistic Regression does perform better than Linear Regression for some predictions, the improvement is small, and Cox Regression performs worse than either Linear Regression or Logistic Regression. (Recall that Cox Regression assumes that hazards are proportional and constant over time; as noted in the Discussion, for at least some patients this is not true.)

**Model development: Personalization and Clusters**
As discussed in the Introduction, because patient (and patient-donor) features are very heterogeneous, the predictive models to be used should be *personalized* to (the features of) a patient. We accomplish this personalization by using the data to identify *predictive clusters* – clusters of patients for whom predictions can best be made by using some specific learner, trained in some specific way (i.e. with coefficients fitted to some specific training set) – and then combining the predictions obtained over the clusters that contain the array of features of the particular patient for whom we need a prediction. This approach allows us to construct a complex predictor from simple learners.

A potential problem with any such construction is that the predictor becomes *too complex* and hence overfits. One way to control model complexity is to append a term that penalizes for complexity – but then the magnitude of the penalty term would become a parameter of the model. Instead, our approach controls model complexity *automatically* by training learners on one part of the development set and then validating on a different part of the development set. We give an outline below; further details, including the pseudocode for the algorithm, can be found in the Supplementary Materials.

**Model development: Tree of predictors**
Our method recursively splits the space $X$ (of patient features or patient-donor-compatibility features) into disjoint subsets (clusters) and, for each cluster, creates a specific predictor by training a learner from the given base class $R$ of learners (Cox Regression, Linear Regression, Logistic Regression) using the development set. This creates a *tree of predictors*: a tree $T$ of clusters (*nodes*) of the feature space $X$ together with a predictor $h_C$ associated to each cluster $C$ in the tree. The overall prediction for a particular patient (or patient-donor pair) having a specific array of features is formed by finding the unique terminal node (cluster) of the tree to which that array of features belongs and the unique path through the tree from the initial node to this terminal node (this is precisely the set of clusters to which the patient belongs) and forming a weighted average of the predictions of the predictors along this path, with weights determined (for the particular path) optimally by linear regression.

The construction of the optimal tree of predictors is recursive; at each stage it builds on the tree previously constructed. We initialize the construction (Stage 0) by defining the initial node of the tree to be the entire space $X$ of features. To determine the predictor $h_X$ to be assigned to the initial node, we train each of the learners in $R$ globally; that is, for each learner we find the coefficients that create the predictor that best fits the development set. Among these predictors, we set $h_X$ to be the one that yields the overall best fit. Having done this we have built a (trivial) tree of predictors (i.e., a single node and predictor). At each successive stage $n+1$ we begin with the tree $T_n$ of predictors constructed in previous stages. For each terminal node $C$ of $T_n$ we proceed as follows. Fix a feature $i$ and a threshold $\tau_i$. (If necessary, we discretize all continuous features.) Define

$$C^- = \{x \in C : x_i < \tau_i\}, \quad C^+ = \{x \in C : x_i \geq \tau_i\}$$

Fix learners $L^-, L^+ \in R$ and nodes $A^-, A^+ \in T_n$ that weakly precede $C^-, C^+$, respectively. Train $L^-$ on $A^-$ (i.e., find the coefficients of $L^-$ that best fit the portion of the development data with features in $A^-$) and train $L^+$ on $A^+$; call the resulting predictors $h^-, h^+$. Using $h^-$ on $C^-$ and $h^+$ on $C^+$ yields a predictor $h^- \cup h^+$ on $C^- \cup C^+ = C$. Among all choices of the feature $i$, the threshold $\tau_i$, the learners $L^-, L^+$ and the training sets $A^-, A^+$, find those that maximize the AUC of $h^- \cup h^+$ on $C$. (See the precise details in the Supplementary Materials.) This defines a split of $C = C^- \cup C^+$ and predictors associated with the sets $C^-, C^+$. If no further improvement is possible, we stop splitting. It can be shown that the process stops after some finite number of stages.

Note that at each stage the construction *jointly* chooses the feature, the threshold, the learner and the training set to *optimize the gain in predictive power* and the construction stops when no further gain is possible; the end result of the process is the *optimal tree of predictors*.

To determine the prediction for a patient having the array $x$ of features, we find the set $\Pi$ of all nodes in the tree to which the array of features $x$ belongs; this is the unique path from the initial node $X$ to the unique terminal node to which $x$ belongs. The overall prediction for $x$ is the weighted average

$$H(x) = \sum_{C \in \Pi} w(C, \Pi) \cdot h_C(x)$$

where the weights $w(C, \Pi)$ are determined optimally by linear regression. (See Supplementary Materials for precise details.)

A typical optimal tree of predictors is shown in Figure 5. (For other prediction trees, see the Supplementary Materials.) Figure 5 shows the feature and threshold used to create each split. Within each cluster, it shows the learner used to create the predictor for that node, the total number of patients within that cluster and the index of the three most relevant features for that cluster. (The list of relevant features is shown in lower part of Figure 5.) Note that, as is typical of our method, the tree constructed is not very deep – the recursive process of construction stops fairly quickly because no further improvement is possible. (See Supplementary Materials for some additional trees.) Consequently, only a few of these clusters are small.

## CONCLUSION

We offer a new method for personalized risk prediction using a novel machine learning technique. Our method is explicitly designed to address the heterogeneity of the patients/donor populations and the fact that the effect of particular features on survival is different for different subpopulations. Moreover, our method is also designed to capture the interaction *between* features and the differences in interactions across different subpopulations. Finally, our method captures the different effects of features and interactions between features for different time horizons. Our method outperforms existing clinical scores and state-of-the-art machine learning methods and is easily interpretable and applicable by clinicians. The present study has important clinical implications for the practice and policy of heart transplantation. The general methodology developed here has wide applicability to the construction of personalized risk scores in other medical domains.

**Supplementary Materials**

- ADDITIONAL DISCUSSIONS

- FIGURES

**Figure S1:** Flow chart from heart failure to post heart transplant

**Figure S2:** Pseudo-codes of ToPs/R

**Figure S3.** Tree of Predictors for the survival in waitlist (3-month) discovered by ToPs/R

**Figure S4.** Tree of Predictors for the survival in waitlist (3-year) discovered by ToPs/R

**Figure S5.** Tree of Predictors for the post-transplantation survival (3-month) discovered by ToPs/R

**Figure S6**. Histograms for predictions in 6 clusters of tree of predictors for the survival in wait-list (3-year)

- TABLES

**Table S1.** Features used in existing medical scores for survival in wait-list with relevance scores (HFSS, SHFM, MAGGIC)

**Table S2.** Features used in existing medical scores for survival after transplant with relevance scores (DRI, RSS, IMPACT)

**Table S3.** Features used in ToPs/R

**Table S4.** Feature used in ToPs/R sorted by relevance score. (Used features selected by CFS are represented in gray)

**Table S5.** Abbreviation

**Acknowledgments:** We are grateful for comments from Prof. Folkert Asselbergs (UMC, the Netherlands), Prof. Paolo Emilio Puddu (La Sapienza, Italy) and Prof. Cesare M N Terracciano (Imperial College, UK). Thank others for any contributions. **Funding**: This work was supported by the Office of Naval Research (ONR) and the NSF (Grant number: ECCS1462245). **Author contributions**: JY, WZ, MvdS contributed equally to all aspects of the paper. AB, MC contributed detailed medical knowledge. AMA contributed research and computing support. **Competing interests**: No competing interests. **Data and materials availability**: All data used in this paper is publicly available in the UNOS archive (available at https://www.unos.org/data/).

# Tables and Figures
# Personalized Survival Predictions for Cardiac Transplantation
# via Trees of Predictors

**- Tables**

**Table 1.** Comparison of AUC values among ToPs/R, existing clinical risk scores, basic regression methods, and machine learning benchmarks for survival prediction in wait-list (3-month, 1-year, 3-year, and 10-year). (*: p-value < 0.01)

**Table 2.** Comparison of AUC values among ToPs/R, existing clinical risk scores, basic regression methods, and machine learning benchmarks for survival prediction post-transplantation (3-month, 1-year, 3-year, and 10-year). (*: p-value < 0.01)

**Table 3.** ToPs/R improvement percentage over other methods for survival prediction in wait-list and post-transplantation

**- Figures**

**Figure 1.** (a) ROC curve for wait-list survival prediction, (b) ROC curve for post-transplantation survival prediction

**Figure 2.** (a) Informative features for each cluster for 3-month wait-list survival prediction, (b) 3-year wait-list survival prediction, (c) for 3-month post-transplantation survival prediction, (d) for 3-year post-transplantation survival prediction, (e) for wait-list survival across the time horizon, (f) for wait-list survival across the time horizon

**Figure 3.** Tree of predictors for 3-year post-transplantation survival

**Figure 4.** (a) Wait-list survival curves for LVAD / LVAD eligible / Entire patients in 2005-2009 and 2010-2015, (b) Post-transplantation survival curves for LVAD / LVAD eligible / Entire patients in 2005-2009 and 2010-2015

**Figure 5.** Tree of predictors for 3-year survival in wait-list

| Algorithms | Risk Score | AUC (95% Confidence Bound) | | | |
|---|---|---|---|---|---|
| | | 3-month survival | 1-year survival | 3-year survival | 10-year survival |
| **Our Model** | ToPs/R | **0.847** (0.846 – 0.848) | **0.813** (0.812 – 0.814) | **0.792** (0.791 – 0.793) | **0.760** (0.759 – 0.761) |
| **Existing Clinical Risk Scores** | HFSS | **0.618** (0.617 – 0.619) | **0.624** (0.623 – 0.625) | **0.621** (0.620 – 0.622) | **0.615** (0.614 – 0.616) |
| | MAGGIC | **0.630** (0.629 - 0.631) | **0.641** (0.640 – 0.642) | **0.643** (0.642 – 0.644) | **0.629** (0.628 – 0.630) |
| | SHFM | **0.601** (0.599 – 0.602) | **0.603** (0.602 – 0.604) | **0.604** (0.603 – 0.605) | **0.610** (0.609 – 0.611) |
| **Our Regressions** | Cox Regression | **0.716** (0.715 – 0.718) | **0.708** (0.707 – 0.708) | **0.672** (0.671 – 0.673) | **0.587** (0.586 – 0.588) |
| | Linear Regression | **0.736** (0.736 – 0.737) | **0.734** (0.733 – 0.736) | **0.719** (0.718 – 0.720) | **0.662** (0.661 – 0.663) |
| | Logistic Regression | **0.760** (0.758 – 0.761) | **0.731** (0.729 – 0.732) | **0.715** (0.713 – 0.716) | **0.705** (0.704 – 0.706) |
| **Machine Learning Benchmarks (Boosting-based)** | AdaBoost | **0.818** (0.816 – 0.820) | **0.787** (0.785 – 0.788) | **0.777** (0.776 – 0.779) | **0.745** (0.744 – 0.746) |
| | Deep Boost | **0.821** (0.820 – 0.822) | **0.790** (0.789 – 0.791) | **0.773** (0.772 – 0.774) | **0.739** (0.738 – 0.740) |
| | Logit Boost | **0.745** (0.744 – 0.746) | **0.737** (0.736 – 0.739) | **0.723** (0.722 – 0.724) | **0.678** (0.677 – 0.679) |
| | XGBoost | **0.824** (0.823 – 0.825) | **0.794** (0.792 – 0.795) | **0.778** (0.777 – 0.779) | **0.746** (0.744 – 0.747) |
| **Machine Learning Benchmarks (Tree-based)** | Decision Tree | **0.819** (0.818 – 0.820) | **0.783** (0.782 – 0.785) | **0.764** (0.763 – 0.766) | **0.744** (0.743 – 0.745) |
| | Random Forest | **0.824** (0.823 – 0.826) | **0.793** (0.791 – 0.794) | **0.774** (0.773 – 0.776) | **0.728** (0.727 – 0.729) |
| **Machine Learning Benchmarks (Others)** | LASSO | **0.759** (0.758 – 0.760) | **0.733** (0.732 – 0.734) | **0.722** (0.721 – 0.724) | **0.686** (0.684 – 0.687) |
| | Neural Networks | **0.788** (0.786 – 0.789) | **0.781** (0.780 – 0.782) | **0.770** (0.769 – 0.771) | **0.741** (0.740 – 0.743) |

**Table 1.** Comparison of AUC values among ToPs/R, existing clinical risk scores, basic regression methods, and machine learning benchmarks for survival prediction in wait-list (3-month, 1-year, 3-year, and 10-year). (p-values < 0.01)

| Algorithms | Risk Score | AUC (95% Confidence Bound) | | | |
|---|---|---|---|---|---|
| | | 3-month survival | 1-year survival | 3-year survival | 10-year survival |
| **Our Model** | ToPs/R | **0.676** (0.675 – 0.678) | **0.664** (0.663 – 0.665) | **0.634** (0.633 – 0.635) | **0.636** (0.635 – 0.637) |
| **Existing Clinical Risk Scores** | DRI | **0.572** (0.571 – 0.573) | **0.562** (0.561 – 0.563) | **0.551** (0.550 – 0.552) | **0.550** (0.549 – 0.551) |
| | IMPACT | **0.581** (0.580 – 0.582) | **0.570** (0.569 – 0.571) | **0.552** (0.551 – 0.554) | **0.531** (0.530 – 0.532) |
| | RSS | **0.563** (0.562 – 0.564) | **0.564** (0.553 – 0.565) | **0.558** (0.557 – 0.559) | **0.567** (0.566 – 0.568) |
| **Our Regressions** | Cox Regression | **0.607** (0.606 – 0.608) | **0.596** (0.595 – 0.597) | **0.573** (0.572 – 0.574) | **0.562** (0.561 – 0.563) |
| | Linear Regression | **0.641** (0.640 – 0.642) | **0.623** (0.622 – 0.624) | **0.603** (0.602 – 0.605) | **0.616** (0.614 – 0.617) |
| | Logistic Regression | **0.638** (0.636 – 0.639) | **0.621** (0.620 – 0.622) | **0.602** (0.601 – 0.603) | **0.616** (0.615 – 0.617) |
| **Machine Learning Benchmarks (Boosting-based)** | AdaBoost | **0.651** (0.649 – 0.652) | **0.630** (0.629 – 0.631) | **0.603** (0.602 – 0.605) | **0.616** (0.614 – 0.617) |
| | Deep Boost | **0.646** (0.645 – 0.648) | **0.635** (0.634 – 0.636) | **0.610** (0.609 – 0.611) | **0.613** (0.612 – 0.615) |
| | Logit Boost | **0.637** (0.636 – 0.638) | **0.622** (0.621 – 0.623) | **0.596** (0.595 – 0.597) | **0.613** (0.612 – 0.614) |
| | XGBoost | **0.618** (0.617 – 0.6200) | **0.608** (0.607 – 0.609) | **0.588** (0.587 – 0.589) | **0.615** (0.614 – 0.616) |
| **Machine Learning Benchmarks (Tree-based)** | Decision Tree | **0.630** (0.629 – 0.631) | **0.611** (0.610 – 0.612) | **0.590** (0.588 – 0.591) | **0.599** (0.598 – 0.600) |
| | Random Forest | **0.653** (0.652 – 0.654) | **0.641** (0.640 – 0.642) | **0.611** (0.610 – 0.613) | **0.619** (0.618 – 0.621) |
| **Machine Learning Benchmarks (Others)** | LASSO | **0.627** (0.626 – 0.629) | **0.615** (0.614 – 0.616) | **0.594** (0.593 – 0.595) | **0.616** (0.615 – 0.617) |
| | Neural Networks | **0.641** (0.640 – 0.643) | **0.638** (0.637 – 0.639) | **0.610** (0.609 – 0.611) | **0.615** (0.613 – 0.616) |

**Table 2.** Comparison of AUC values among ToPs/R, existing clinical risk scores, basic regression methods, and machine learning benchmarks for survival prediction post-transplantation (3-month, 1-year, 3-year, and 10-year). (p-values < 0.01)

|  | Algorithms | Risk Score | ToPs/R % improvement ||||
|---|---|---|---|---|---|---|
|  |  |  | 3-month survival | 1-year survival | 3-year survival | 10-year survival |
| **Survival prediction in Wait-list** | **Our Model** | **ToPs/R** |  |  |  |  |
|  | **Existing Clinical Risk Scores** | HFSS | 59.9% | 50.2% | 45.1% | 37.6% |
|  |  | MAGGIC | 58.6% | 47.9% | 41.8% | 35.2% |
|  |  | SHFM | 61.6% | 52.9% | 47.5% | 38.4% |
|  | **Our Regressions** | Cox Regression | 46.0% | 36.1% | 36.6% | 41.9% |
|  |  | Linear Regression | 42.0% | 29.6% | 26.1% | 28.9% |
|  |  | Logistic Regression | 36.3% | 30.6% | 27.1% | 18.6% |
|  | **Machine Learning Benchmarks** — Boosting-based | AdaBoost | 15.8% | 12.4% | 6.6% | 5.7% |
|  |  | Deep Boost | 14.3% | 11.0% | 8.4% | 7.9% |
|  |  | Logit Boost | 39.9% | 28.9% | 24.9% | 25.5% |
|  |  | XGBoost | 13.0% | 9.5% | 6.4% | 5.5% |
|  | Tree-based | Decision Tree | 15.4% | 13.7% | 11.8% | 6.1% |
|  |  | Random Forest | 12.8% | 9.8% | 7.8% | 11.6% |
|  | Others | LASSO | 36.3% | 29.8% | 25.1% | 23.5% |
|  |  | Neural Networks | 27.8% | 14.5% | 9.7% | 7.1% |
| **Survival prediction in Post-transplant** | **Our Model** | **ToPs/R** |  |  |  |  |
|  | **Existing Clinical Risk Scores** | DRI | 24.4% | 23.3% | 18.5% | 19.2% |
|  |  | IMPACT | 22.8% | 21.8% | 18.2% | 22.5% |
|  |  | RSS | 26.0% | 22.8% | 17.2% | 15.9% |
|  | **Our Regressions** | Cox Regression | 17.6% | 16.8% | 14.2% | 16.9% |
|  |  | Linear Regression | 9.9% | 10.7% | 7.7% | 5.4% |
|  |  | Logistic Regression | 10.7% | 11.4% | 7.9% | 5.2% |
|  | **Machine Learning Benchmarks** — Boosting-based | AdaBoost | 7.4% | 9.1% | 8.0% | 4.9% |
|  |  | Deep Boost | 8.5% | 7.9% | 6.1% | 5.9% |
|  |  | Logit Boost | 10.8% | 11.1% | 9.3% | 6.0% |
|  |  | XGBoost | 15.2% | 14.1% | 11.2% | 5.5% |
|  | Tree-based | Decision Tree | 12.6% | 13.6% | 10.8% | 9.3% |
|  |  | Random Forest | 6.7% | 6.2% | 5.8% | 4.4% |
|  | Others | LASSO | 13.1% | 12.7% | 9.7% | 5.3% |
|  |  | Neural Networks | 9.8% | 7.2% | 6.1% | 5.6% |

**Table 3.** ToPs/R improvement percentage over other methods for survival prediction in wait-list and post-transplantation

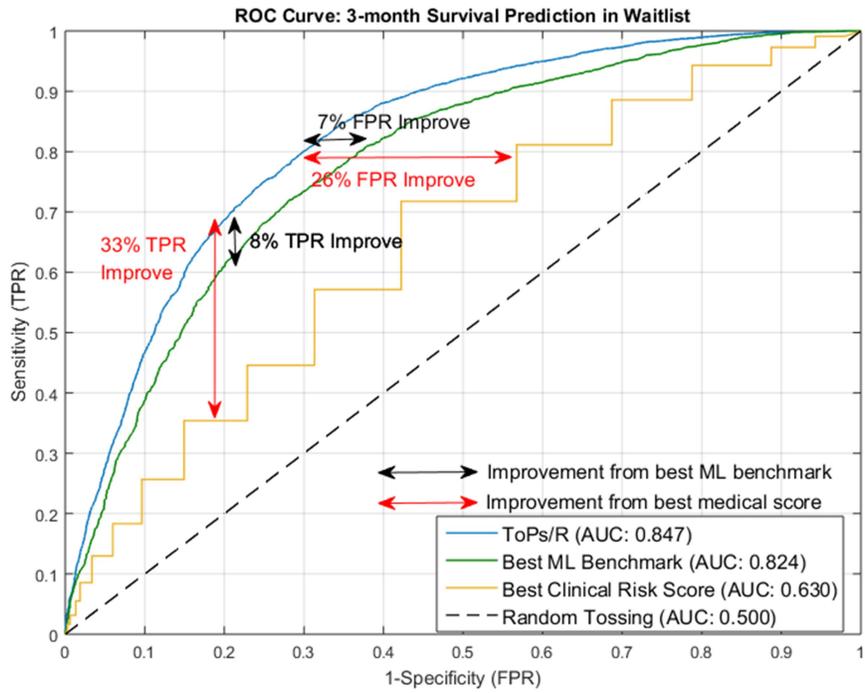

**Figure 1** (a)

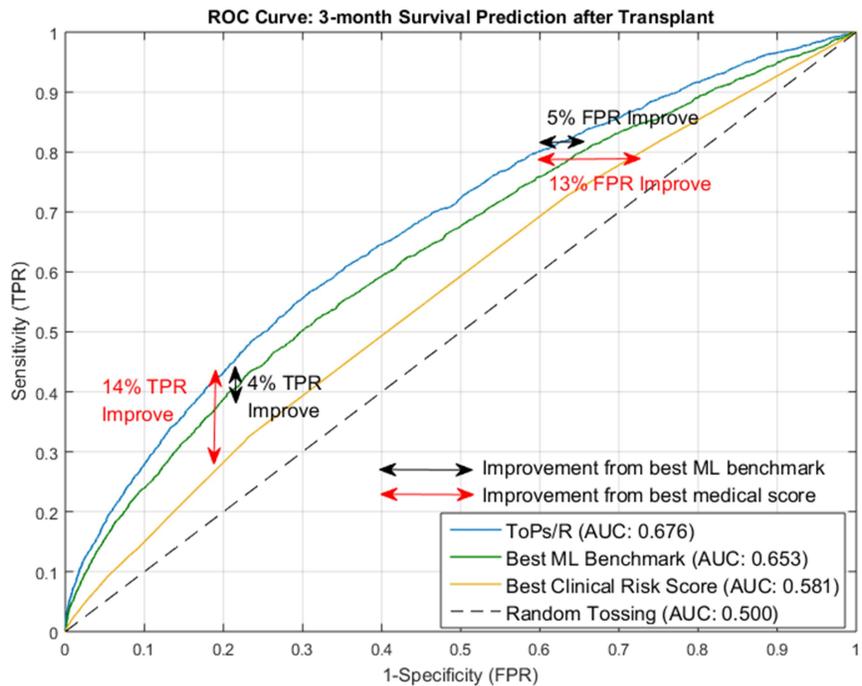

**Figure 1** (b)

**Figure 1.** (a) ROC curve for wait-list survival prediction, (b) ROC curve for post-transplantation survival prediction

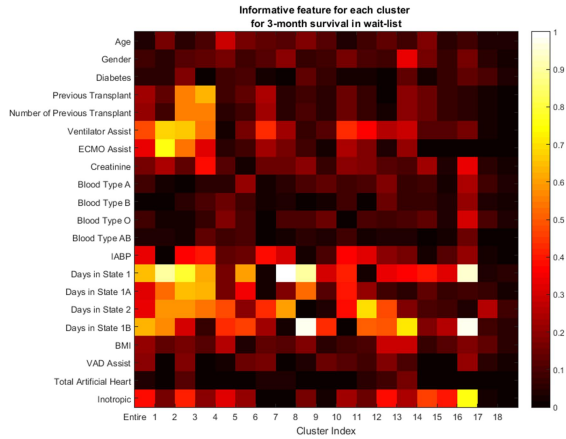
**Figure 2** (a)

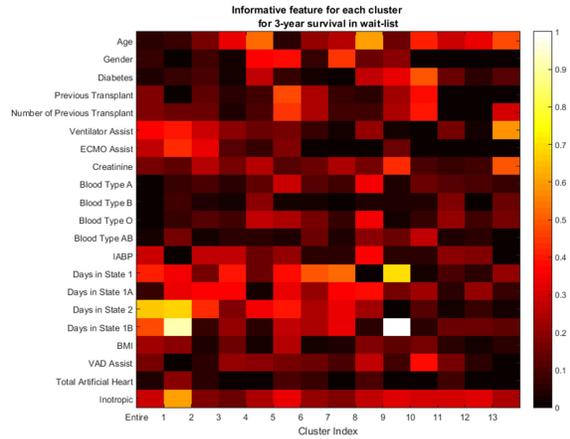
**Figure 2** (b)

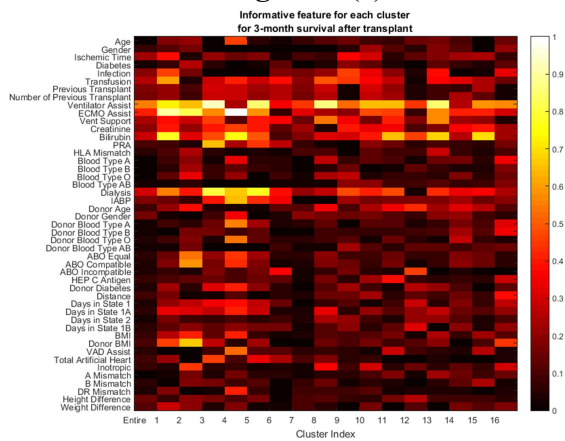
**Figure 2** (c)

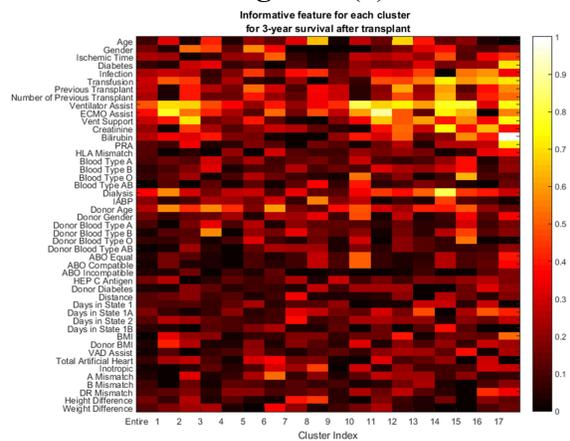
**Figure 2** (d)

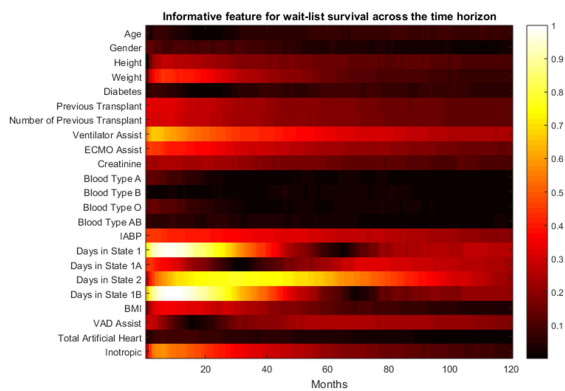
**Figure 2** (e)

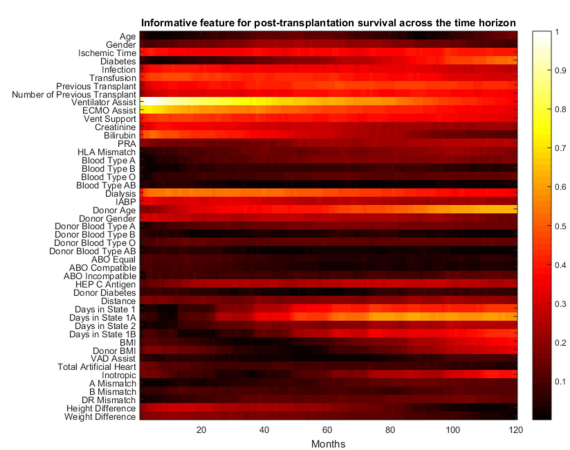
**Figure 2** (f)

**Figure 2.** (a) Informative features for each cluster for 3-month wait-list survival prediction, (b) 3-year wait-list survival prediction, (c) for 3-month post-transplantation survival prediction, (d) for 3-year post-transplantation survival prediction, (e) for wait-list survival across the time horizon, (f) for wait-list survival across the time horizon

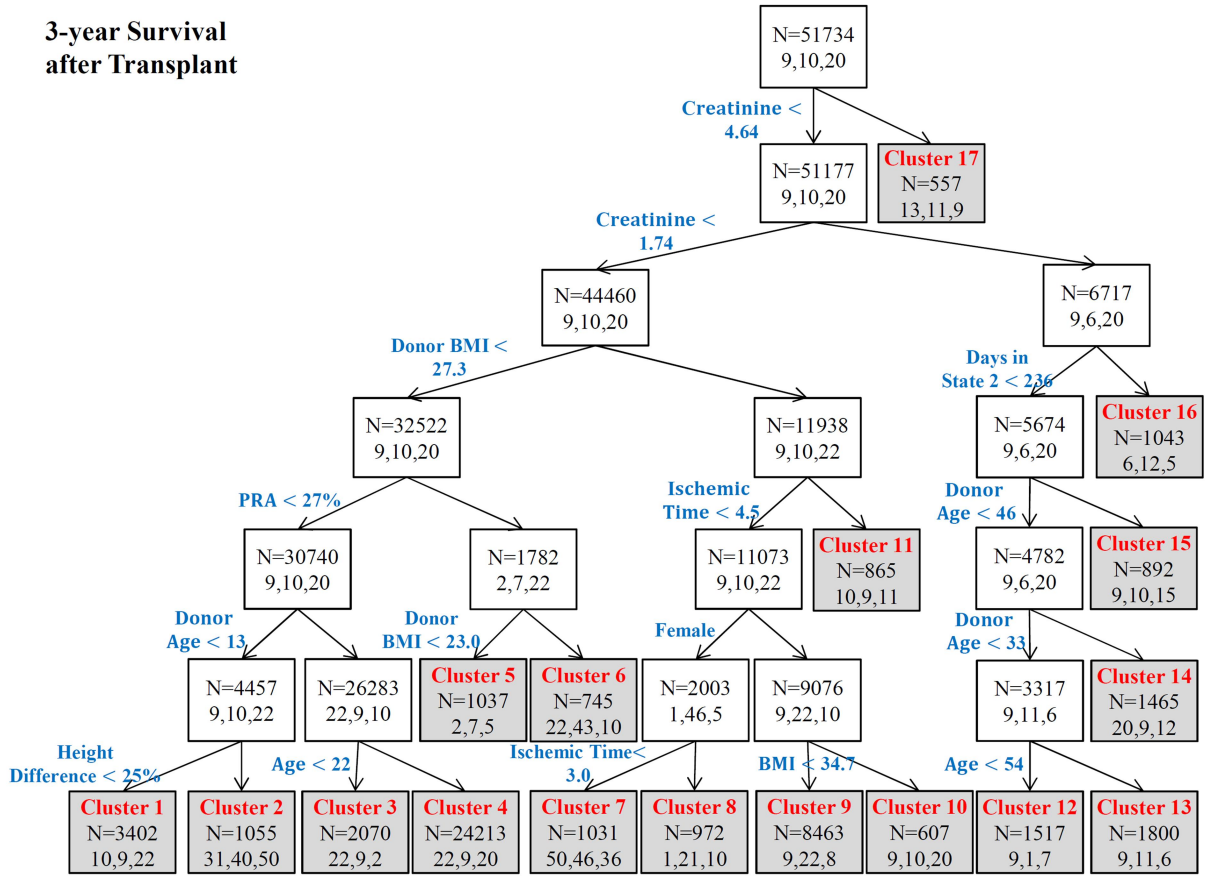

**Figure 3.** Tree of predictors for 3-year post transplantation survival (End nodes shaded gray. 3 most relevant features shown for each node.)

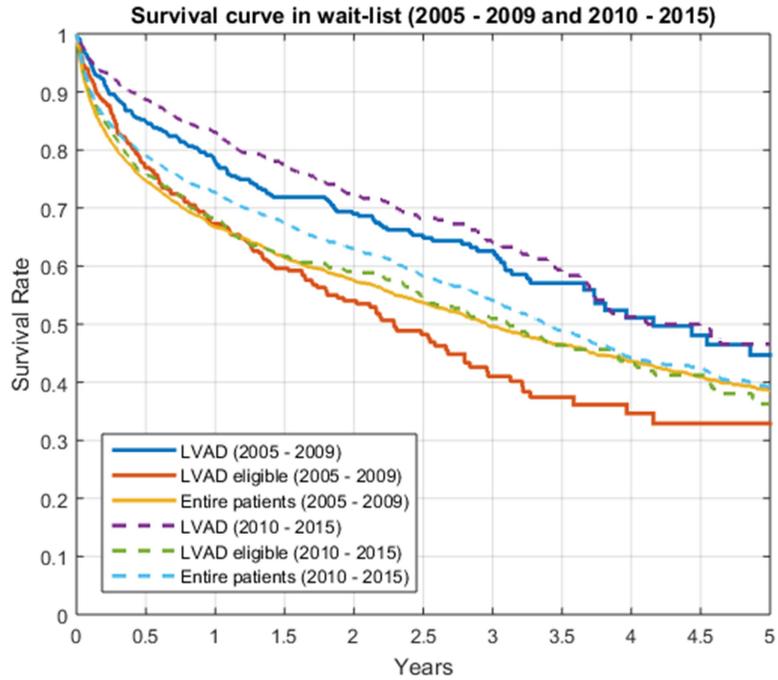

**Figure 4** (a)

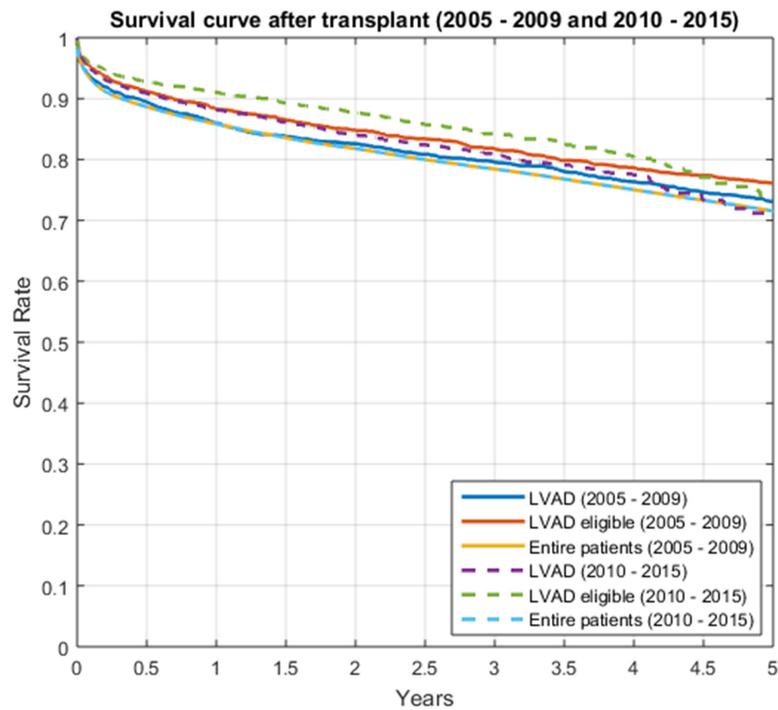

**Figure 4** (b)

**Figure 4.** (a) Wait-list survival curves for LVAD / LVAD eligible / Entire patients in 2005-2009 and 2010-2015, (b) Post-transplantation survival curves for LVAD / LVAD eligible / Entire patients in 2005-2009 and 2010-2015

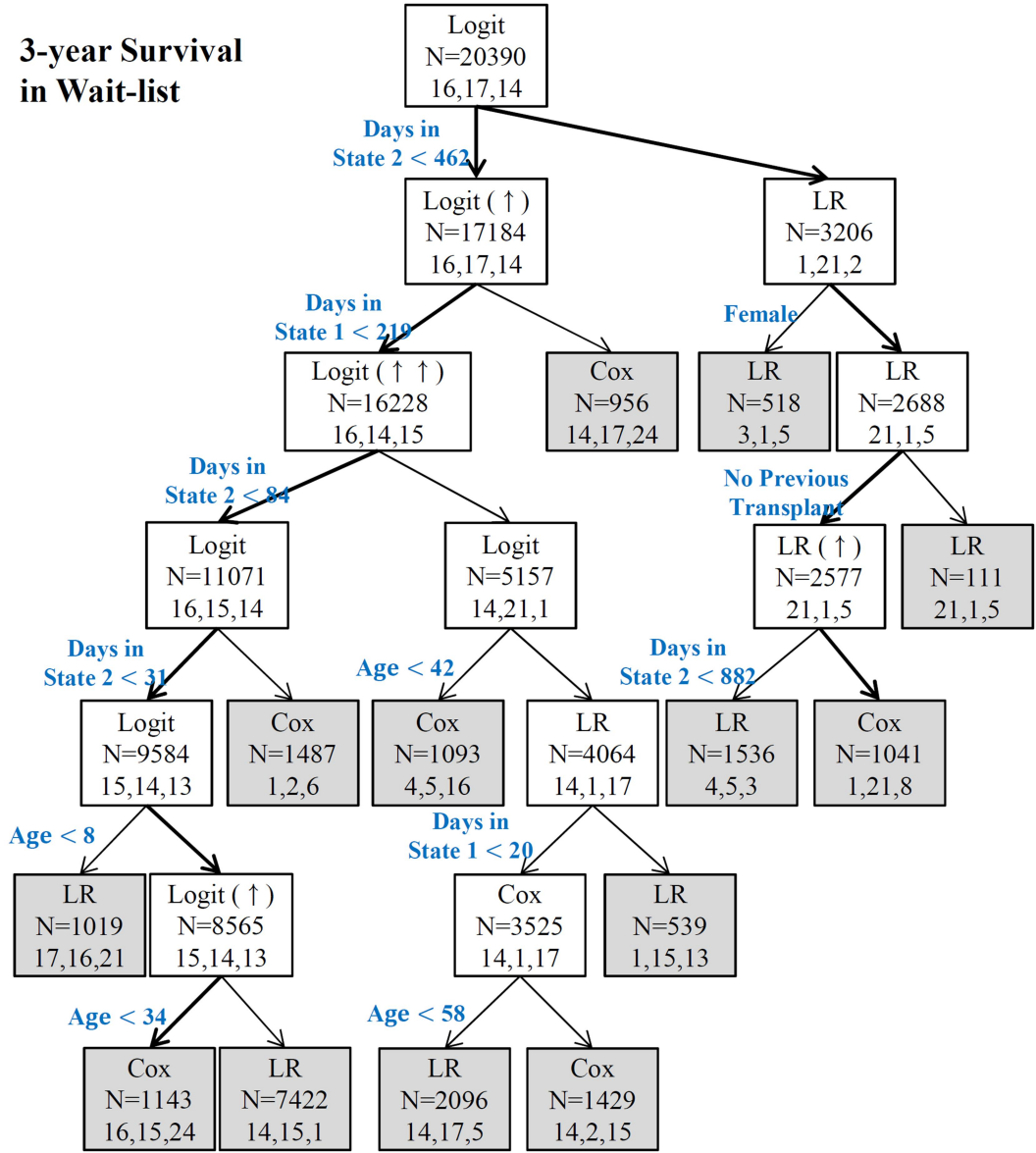

**Figure 5.** Tree of predictors for 3-year survival in wait-list ((End nodes shaded gray. 3 most relevant features shown for each node.)

# Supplementary Materials
# Personalized Survival Predictions for Cardiac Transplantation
# via Trees of Predictors

**- ADDITIONAL DISCUSSIONS**

**- FIGURES**

**Figure S1:** Flow chart from heart failure to post heart transplantation

**Figure S2:** Pseudo-codes of ToPs/R

**Figure S3.** Tree of Predictors for the survival in wait-list (3-month) discovered by ToPs/R

**Figure S4.** Tree of Predictors for the survival in wait-list (3-year) discovered by ToPs/R

**Figure S5.** Tree of Predictors for the post-transplantation survival (3-month) discovered by ToPs/R

**Figure S6**. Histograms for predictions in 6 clusters of tree of predictors for the survival in wait-list (3-year)

**- TABLES**

**Table S1.** Features used in existing medical scores for survival in wait-list with relevance scores (HFSS, SHFM, MAGGIC)

**Table S2.** Features used in existing medical scores for survival after transplantation with relevance scores (DRI, RSS, IMPACT)

**Table S3.** Features used in ToPs/R

**Table S4.** Feature used in ToPs/R sorted by relevance score. (Features selected by CFS are represented in gray)

**Table S5.** Abbreviations

# ADDITIONAL DISCUSSIONS

**Predictive features and discriminative features**

Our method successively splits the population into subpopulations by finding features that *discriminate* between subpopulations for which the best-fitting predictive models are different. A separate, tailored predictive model is then constructed for each given subpopulation using the features that are *predictive* of survival for that subpopulation. *Discriminative* and *predictive* features need not be the same. The following analogy may illustrate the distinction between discrimination and prediction. Assume that we have a population of patients for whom we want to perform accurate diagnosis for some underlying disease, and assume that each patient can undergo one of two possible diagnostic tests: a blood test and a genetic test. Suppose we know that for a left-handed patient, the genetic test is more accurate than the blood test, whereas for a right-handed patient, the blood test has a higher diagnostic accuracy. In this hypothetical scenario, it is optimal to split the population into a left-handed subpopulation, for which a genetic test is recommended, and a right-handed subpopulation, for which a blood test is recommended. In this example, handedness is a *discriminative* feature: it need not tell us anything about the *true* diagnosis in either subpopulation, but it informs us as to which diagnostic test should be selected for which patient. The diagnostic tests, which are analogous to *predictive models*, depend on different predictive features for the different subpopulations: for the left-handed subpopulation, the "predictive features" are genetic information, whereas for the right-handed subpopulation, the predictive features are hematologic information. In a very similar manner, ToPs uses discriminative features to determine the *right predictive model* for the *right subpopulation*, and uses the predictive features to cast a predictive model for every subpopulation. The identification of a feature as being discriminative or predictive (or both) is driven completely by the data. (As can be seen in the Supplementary Materials, BMI has substantial discriminatory power but little predictive power, ECMO has substantial predictive power but little discriminatory power, and Age has both substantial discriminatory power and substantial predictive power.)

**Complex models versus simple models**
As a general principle, complex models are better able to fit the data than are simple models, but complex models are more likely to suffer from over-fitting than are simple models. Hence it may be better to prefer a simple model (e.g. Linear Regression) rather than a more complex model (e.g. Logistic Regression) when the data set is "small." In our setting, the entire data set is large but some clusters may be "small"; our method takes into account the size of the cluster and *uses the data* to decide how to resolve the tension between complexity and simplicity.

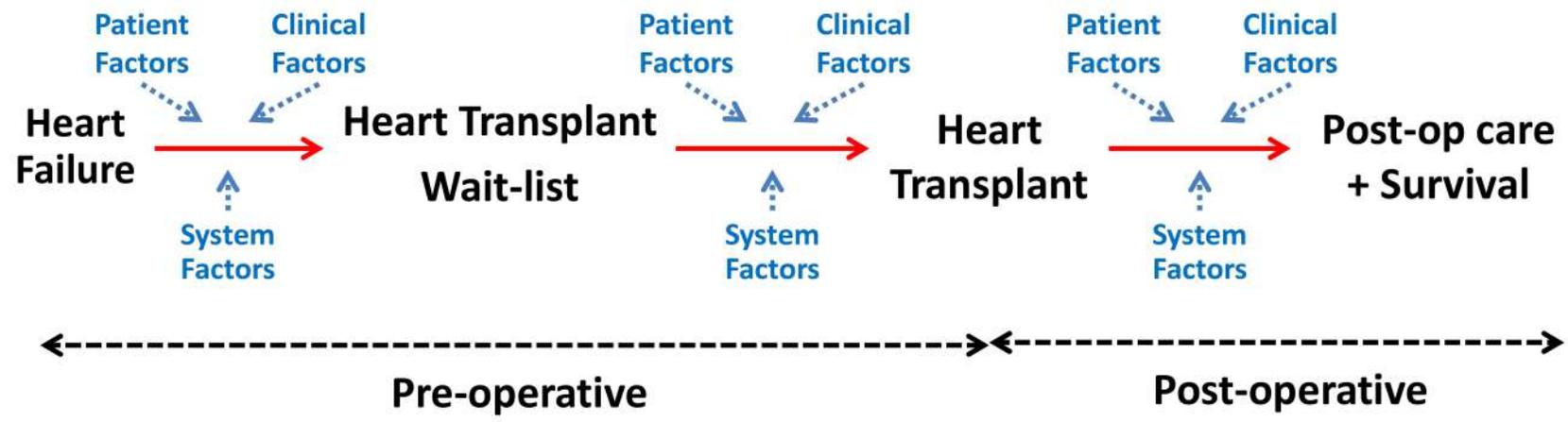

**Figure S1.** Flow chart from heart failure to post heart transplantation

| **Off-line Stage 0: Dividing the dataset ($D$)** |
|---|
| **Input:** Entire dataset $D$ |
| **Divide** $D$ into disjoint training set ($S$), the first validation set ($V^1$), the second validation set ($V^2$) and testing set ($T$) which satisfy $D = S \cup V^1 \cup V^2 \cup T$ |
| **Output:** Training set $S$, validation sets $V^1$, $V^2$, and testing set $T$ |
| **Off-line Stage 1: Growing the Optimal Tree of Predictors** |
| **Input:** Feature space $X$, a set of algorithms $A$, training set $S$, the first validation set $V^1$ |
| **First step:** Initial tree of predictors $= (X, h_X)$ where $h_X = \arg\min_{A \in A} L(A(S), V^1)$ |
| **Recursive step:** |
| **Input:** Current tree of predictors $(T, \{h_C\})$ |
| **For** each terminal node $C \in T$ **do** |
|    **For** a feature $i$ and a threshold $\tau_i$ **do** |
|      Set $C^-(\tau_i) = \{x \in C : x_i < \tau_i\}$, $C^+(\tau_i) = \{x \in C : x_i \geq \tau_i\}$ Then, |
|      $\{i^*, \tau_i^*, h_{C^-(\tau_i^*)}, h_{C^+(\tau_i^*)}\} = \arg\min L(h^- \cup h^+, V^1(C^-(\tau_i)) \cup V^1(C^+(\tau_i)))$ |
|      where $h^- \in A(C^-(\tau_i)^\uparrow)$, $h^+ \in A(C^+(\tau_i)^\uparrow)$ |
|    **End for** |
| **End for** |
| **Stopping Criteria:** $L(h_C, V^1(C)) \leq \min L(h^- \cup h^+, V^1(C^-(\tau_i)) \cup V^1(C^+(\tau_i)))$ |
| **Output:** Locally optimal tree of predictors $(T, \{h_C\})$ |
| **Off-line Stage 2: Weights Optimization on the Path** |
| **Input:** Locally optimal tree of predictors $(T, \{h_C\})$, the second validation set $V^2$ |
| **For** each terminal node $\overline{C}$ and the corresponding path $\Pi$ from $X$ to $\overline{C}$ |
|    For each weight vector $w = (w_C)$, |
|    Define $H_w = \sum_{C \in \Pi} w_C h_C$, Then, |
|    $w^*(\Pi) = (w^*(\Pi, C)) = \arg\min L(H_w, V^2(\overline{C}))$ |
| **End for** |
| **Output:** Optimized weights $w^*(\Pi)$ for each terminal node $\overline{C}$ and corresponding path $\Pi$ |
| **On-line Stage: Overall Predictor** |
| **Input:** Locally optimal tree of predictors $(T, \{h_C\})$, optimized weights $w^*(\Pi)$, and testing set $T$ |
| Given a feature vector $x$, |
| Find the unique path $\Pi(x)$ from $X$ to terminal node containing $x$; |
| Then, $H(x) = \sum_{C \in \Pi(x)} w^*(\Pi, C) h_C(x)$ |
| **Output:** The final prediction $H(x)$ |

**Figure S2.** Pseudo-codes of ToPs/R

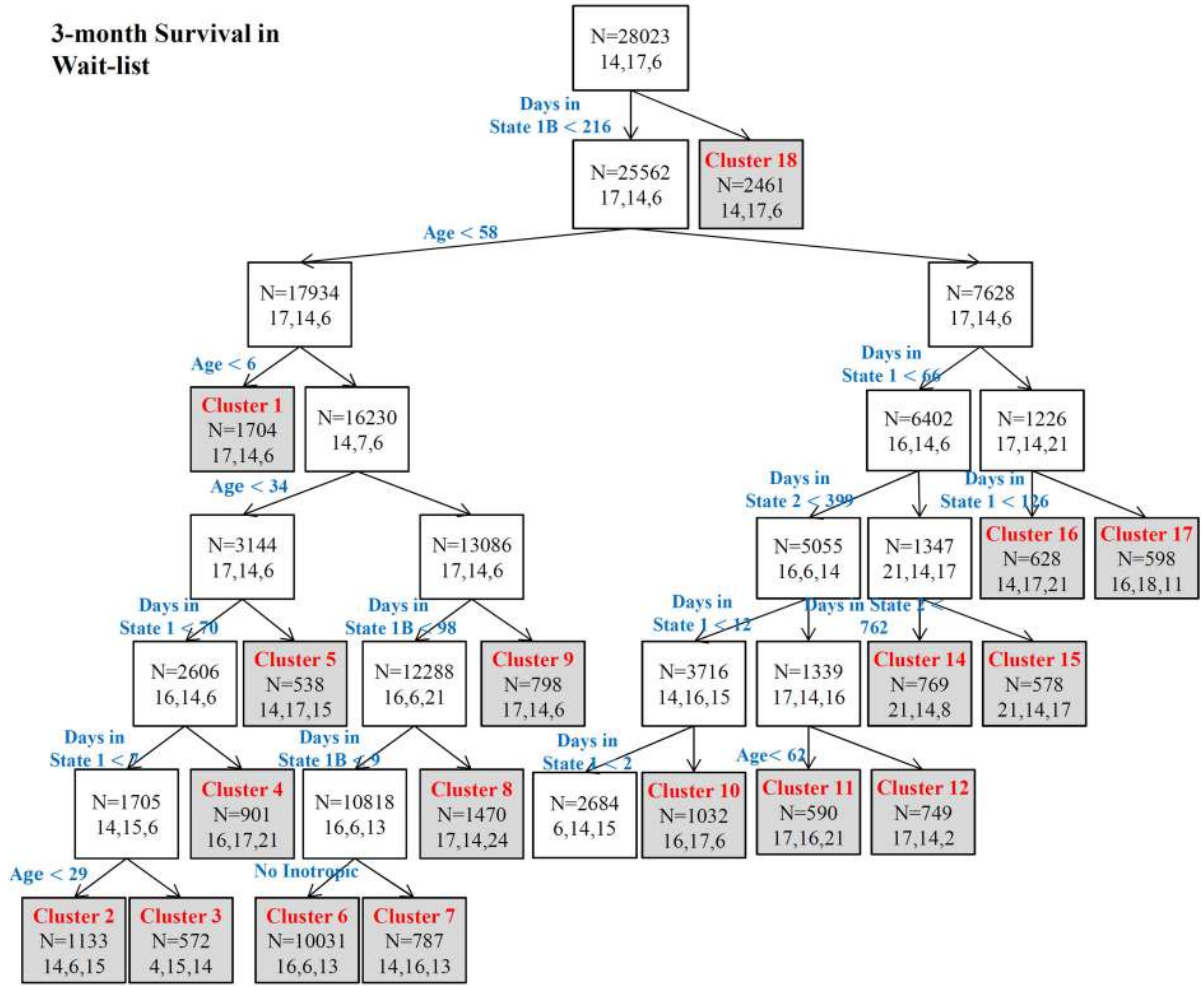

**Figure S3.** Tree of Predictors for the survival in wait-list (3-month) discovered by ToPs/R.

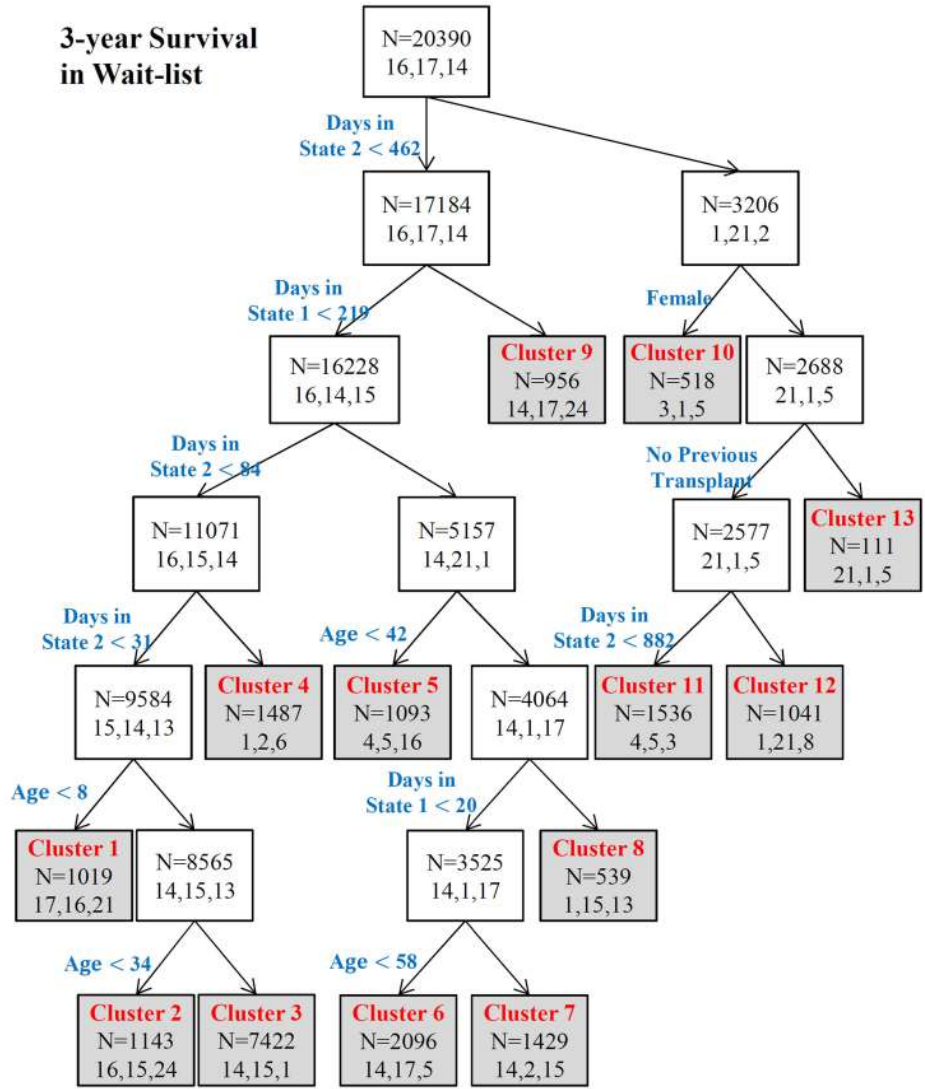

**Figure S4.** Tree of Predictors for the survival in wait-list (3-year) discovered by ToPs/R

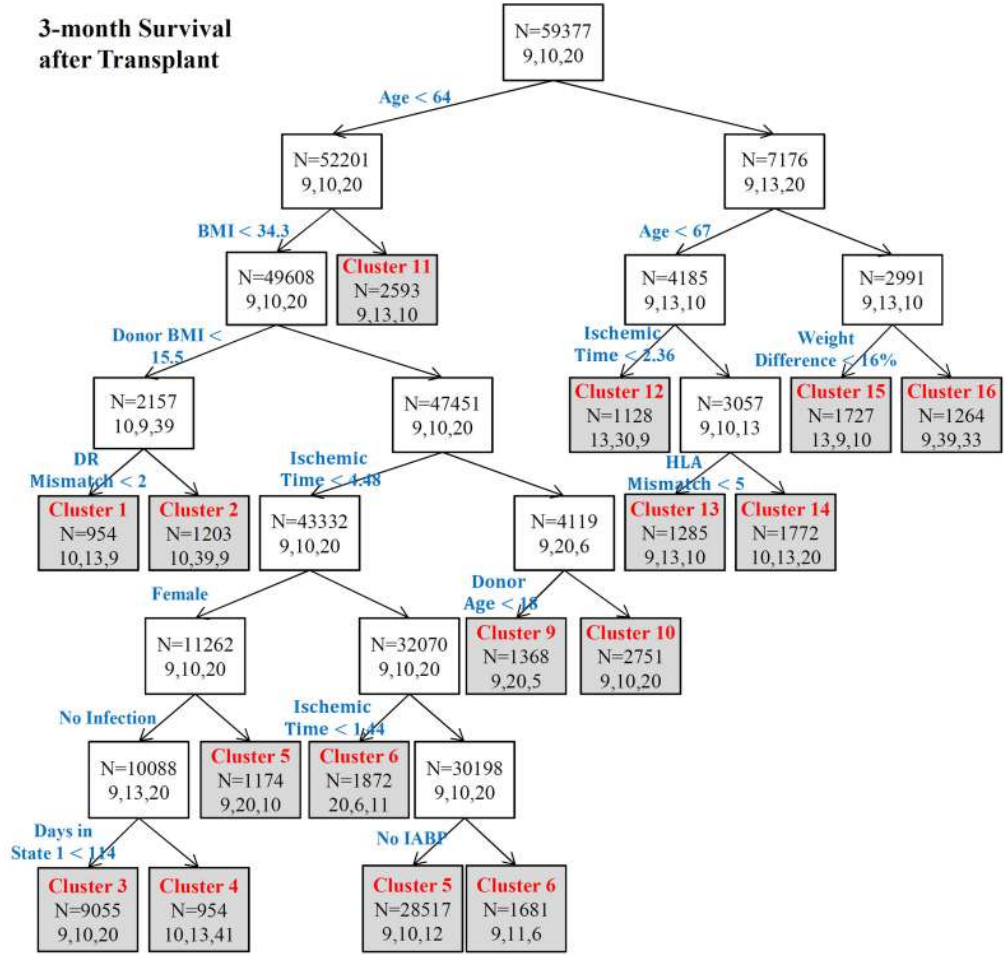

**Figure S5.** Tree of Predictors for the post-transplantation survival (3-month) discovered by ToPs/R

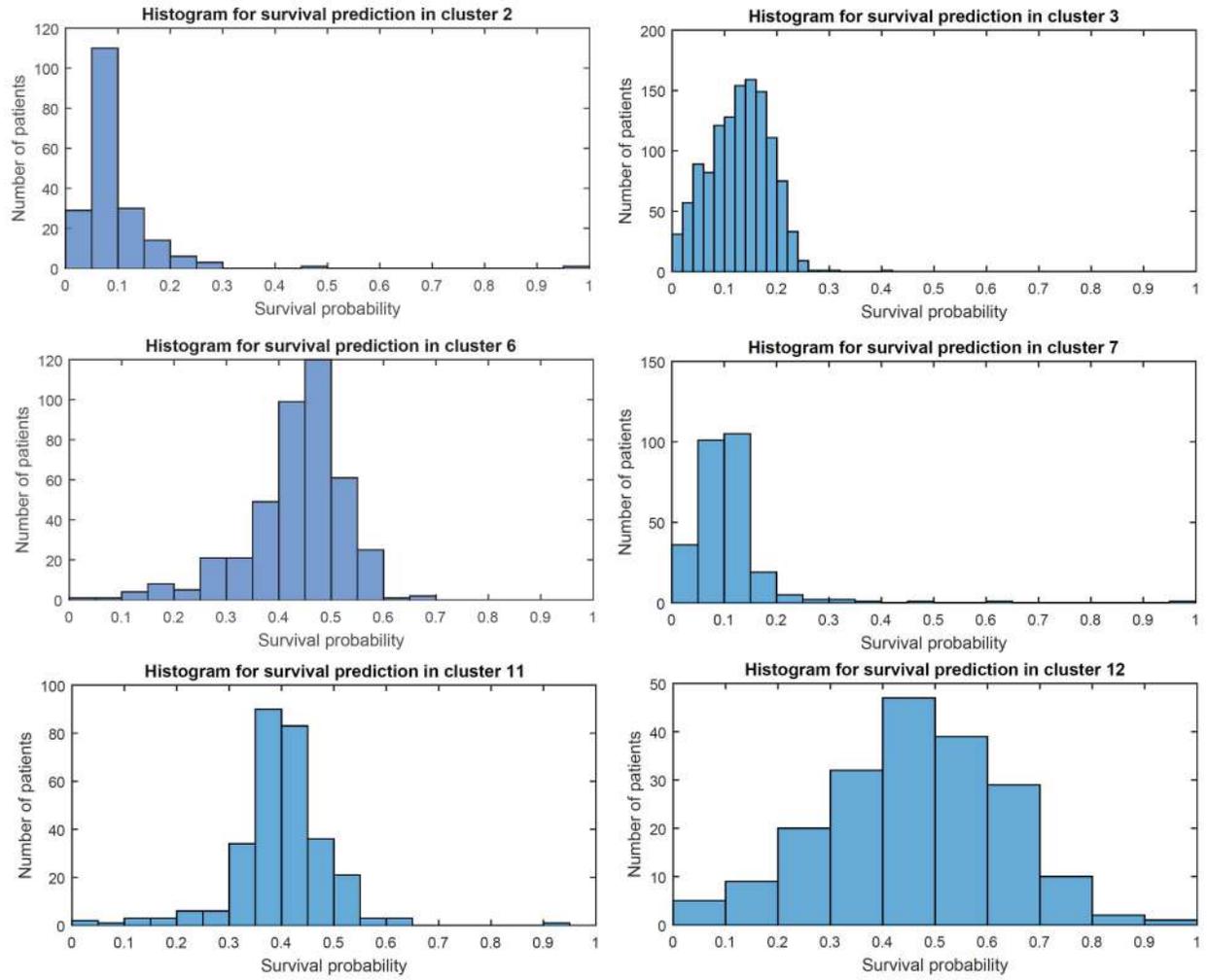

**Figure S6**. Histograms for predictions in 6 clusters of tree of predictors for the survival in wait-list (3-year)

|  | HFSS | | | SHFM | | | MAGGIC | |
| --- | --- | --- | --- | --- | --- | --- | --- | --- |
| No | Feature Name | Relevance Score | No | Feature Name | Relevance Score | No | Feature Name | Relevance Score |
| 1 | LVEF (%) | 0.043 | 1 | Recipient Age (years) | 0.528 | 1 | EF (%) | 0.043 |
| 2 | Mean Blood Pressure (mmHg) | 0.171 | 2 | Recipient Gender | 0.226 | 2 | Recipient Age (Year) | 0.528 |
| 3 | pkVO2 (mg/dL) | 0.098 | 3 | NYHA Functional Classification | 0.173 | 3 | Systolic Blood Pressure | 0.171 |
| 4 | Serum Sodium (mmol/L) |  | 4 | EF (%) | 0.043 | 4 | BMI | 0.361 |
| 5 | Ischemic Cardiomyopathy | 0.213 | 5 | Etiology (Ischemic) | 0.213 | 5 | Creatinine (mmol/L) | 0.182 |
| 6 | Intraventricular Conduction Delay (ms) |  | 6 | Systolic Blood Pressure (mmHg) | 0.171 | 6 | NYHA Class | 0.173 |
| 7 | Resting Heart Rate (bpm) |  | 7 | ACE inhibitor | 0.558 | 7 | Recipient Gender | 0.226 |
|  |  |  | 8 | Diuretic (mg/day) |  | 8 | Current Smoker |  |
|  |  |  | 9 | Recipient Weight (kg) | 0.123 | 9 | Diabetic | 0.595 |
|  |  |  | 10 | Sodium (mEq/L) |  | 10 | Diagnosis of COPD |  |
|  |  |  | 11 | Hemoglobin (g/dL) |  | 11 | First Diagnosis of Heart Failure |  |
|  |  |  | 12 | Lymphocytes (%) |  | 12 | Not on Beta Blocker | 0.151 |
|  |  |  | 13 | Uric Acid (mg/dL) |  | 13 | Not on ACEI/ARB | 0.558 |
|  |  |  | 14 | Cholesterol (mg/dL) |  |  |  |  |

**Table S1.** Features used in existing medical scores for survival in wait-list with relevance scores (HFSS, SHFM, MAGGIC)

|  | DRI | | | RSS | | | IMPACT | |
|---|---|---|---|---|---|---|---|---|
| No | Feature Name | Relevance Score | No | Feature Name | Relevance Score | No | Feature Name | Relevance Score |
| 1 | Ischemic Time (hours) | 0.494 | 1 | Recipient Age (years) | 0.348 | 1 | Recipient Age (years) | 0.348 |
| 2 | Donor Age (years) | 1.000 | 2 | Previous Cardiac Surgery | 0.534 | 2 | Total Bilirubin (mg/dL) | 0.023 |
| 3 | Recipient Race | 0.017 | 3 | Etiology *(Congenital / Amyloidosis)* |  | 3 | Creatinine (mg/dL) | 0.219 |
| 4 | Donor Race | 0.013 | 4 | Diabetes complicated by CVA | 0.874 | 4 | Dialysis between listing and transplant | 0.374 |
| 5 | BUN (mg/dL) | 0.127 | 5 | eGFR (mL/minute) |  | 5 | Recipient Gender | 0.049 |
| 6 | Creatinine (mg/dL) | 0.219 | 6 | Total Bilirubin (mg/dL) | 0.023 | 6 | Total Bilirubin (mg/dL) | 0.023 |
|  |  |  | 7 | Acuity *(Intubated / Hospitalized)* |  | 7 | Heart Failure Etiology |  |
|  |  |  | 8 | RVAD only Support | 0.093 | 8 | Infection | 0.281 |
|  |  |  | 9 | ECMO Support | 0.385 | 9 | IABP Support | 0.285 |
|  |  |  | 10 | Extracorporeal LVAD Support | 0.093 | 10 | Mechanical Ventilator Support | 0.362 |
|  |  |  | 11 | Total Artificial Heart Support | 0.04 | 11 | Recipient Race | 0.017 |
|  |  |  | 12 | Paracorporeal LVAD Support | 0.093 |  |  |  |
|  |  |  | 13 | Donor HEP C | 0.392 |  |  |  |
|  |  |  | 14 | Insulin Dependent Donor | 0.06 |  |  |  |
|  |  |  | 15 | Donor Age (years) | 1.000 |  |  |  |
|  |  |  | 16 | Ischemic Time (hours) | 0.494 |  |  |  |
|  |  |  | 17 | Donor Gender | 0.199 |  |  |  |
|  |  |  | 18 | Recipient Gender | 0.049 |  |  |  |

**Table S2.** Features used in existing medical scores for survival post-transplantation with relevance scores (DRI, RSS, IMPACT)

|  | Recipient features | | | | Donor features | | Donor/Recipient Compatibility features |
|---|---|---|---|---|---|---|---|
| No | Feature Name | No | Feature Name | No | Feature Name | No | Feature Name |
| 1 | Age (years) | 19 | Blood Type AB | 1 | Donor Age (years) | 1 | ABO Compatibility |
| 2 | Gender | 20 | Dialysis | 2 | Donor Gender | 2 | Ischemic Time (hours) |
| 3 | Height (cm) | 21 | IABP | 3 | Donor Height (cm) | 3 | HLA Mismatch |
| 4 | Weight (kg) | 22 | Days in State 1 (days) | 4 | Donor Weight (kg) | 4 | A Mismatch |
| 5 | Diabetes | 23 | Days in State 1A (days) | 5 | Donor Blood Type A | 5 | B Mismatch |
| 6 | Infection | 24 | Days in State 2 (days) | 6 | Donor Blood Type B | 6 | DR Mismatch |
| 7 | Transfusion | 25 | Days in State 1B (days) | 7 | Donor Blood Type O | | |
| 8 | Previous Transplant | 26 | BMI (kg/m$^2$) | 8 | Donor Blood Type AB | | |
| 9 | # of Previous Transplant | 27 | LVAD Assist | 9 | HEP C Antigen | | |
| 10 | Ventilator Assist | 28 | Total Artificial Heart | 10 | Donor Diabetes | | |
| 11 | ECMO Assist | 29 | Inotropic | 11 | Donor BMI (kg/m$^2$) | | |
| 12 | Circulatory Support | 30 | EF (%) | 12 | Donor Infection | | |
| 13 | Creatinine (mg/dL) | 31 | pkVO2 (mg/dL) | 13 | Bilirubin (mg/dL) | | |
| 14 | Bilirubin (mg/dL) | 32 | ACE inhibitor | 14 | Creatinine (mg/dL) | | |
| 15 | PRA (%) | 33 | Beta Blocker | | | | |
| 16 | Blood Type A | | | | | | |
| 17 | Blood Type B | | | | | | |
| 18 | Blood Type O | | | | | | |

**Table S3.** Features used in ToPs

| | Survival in Wait-list | | | Survival after Transplant | |
|---|---|---|---|---|---|
| **No** | **Feature** | **Relevance Score** | **No** | **Feature** | **Relevance Score** |
| 1 | 'Previous Transplant' | 1.000 | 1 | Donor Age | 1.000 |
| 2 | 'Days in State 1' | 0.835 | 2 | Diabetes | 0.874 |
| 3 | '# of Previous Transplant' | 0.799 | 3 | BMI | 0.700 |
| 4 | 'Days in State 1B' | 0.728 | 4 | Days in State 1A | 0.682 |
| 5 | Circulatory Support | 0.717 | 5 | Donor BMI | 0.602 |
| 6 | 'IABP' | 0.696 | 6 | Inotropic | 0.565 |
| 7 | 'Diabetes' | 0.595 | 7 | Donor Infection | 0.545 |
| 8 | 'Blood Type O' | 0.590 | 8 | Days in State 1B | 0.544 |
| 9 | 'Days in State 1A' | 0.582 | 9 | Previous Transplant | 0.534 |
| 10 | ACE-inhibitor | 0.558 | 10 | Days in State 1 | 0.526 |
| 11 | 'ECMO Assist' | 0.556 | 11 | Donor Creatinine | 0.502 |
| 12 | 'Days in State 2' | 0.545 | 12 | Ischemic Time | 0.494 |
| 13 | 'Age' | 0.528 | 13 | # of Previous Transplant | 0.427 |
| 14 | 'Ventilator Assist' | 0.501 | 14 | PRA | 0.394 |
| 15 | Infection | 0.405 | 15 | HEP C Antigen | 0.392 |
| 16 | 'BMI' | 0.361 | 16 | ECMO Assist | 0.385 |
| 17 | PRA | 0.248 | 17 | Dialysis | 0.374 |
| 18 | 'Gender' | 0.226 | 18 | Ventilator Assist | 0.362 |
| 19 | Transfusion | 0.189 | 19 | Age | 0.348 |
| 20 | 'Creatinine' | 0.182 | 20 | Days in State 2 | 0.346 |
| 21 | 'LVAD Assist' | 0.174 | 21 | Donor Height | 0.287 |
| 22 | Beta Blocker | 0.151 | 22 | Transfusion | 0.285 |
| 23 | Dialysis | 0.143 | 23 | IABP | 0.285 |
| 24 | Total Artificial Heart | 0.129 | 24 | Infection | 0.281 |
| 25 | Weight | 0.123 | 25 | Vent Support | 0.27 |
| 26 | Inotropic | 0.121 | 26 | Blood Type A | 0.239 |
| 27 | pkVO2 | 0.098 | 27 | Height | 0.231 |
| 28 | EF | 0.043 | 28 | HLA Mismatch | 0.23 |
| 29 | 'Blood Type AB' | 0.015 | 29 | Creatinine | 0.219 |
| 30 | Bilirubin | 0.012 | 30 | DR Mismatch | 0.21 |
| 31 | 'Blood Type B' | 0.009 | 31 | Blood Type O | 0.206 |
| 32 | 'Height' | 0.004 | 32 | Distance | 0.204 |
| 33 | 'Blood Type A' | 0.000 | 33 | EF(%) | 0.199 |
| | | | 34 | Donor Gender | 0.199 |
| | | | 35 | Donor Blood Type A | 0.177 |
| | | | 36 | A Mismatch | 0.157 |
| | | | 37 | Donor Blood Type O | 0.149 |
| | | | 38 | pkVO2 | 0.132 |
| | | | 39 | Weight | 0.102 |
| | | | 40 | ACE inhibitor | 0.094 |
| | | | 41 | LVAD Assist | 0.093 |
| | | | 42 | B Mismatch | 0.084 |
| | | | 43 | Donor Diabetes | 0.06 |
| | | | 44 | Gender | 0.049 |
| | | | 45 | Blood Type AB | 0.042 |
| | | | 46 | Total Artificial Heart | 0.04 |
| | | | 47 | Donor Blood Type AB | 0.034 |
| | | | 48 | Bilirubin | 0.023 |
| | | | 49 | Blood Type B | 0.021 |
| | | | 50 | ABO_Compatible | 0.016 |
| | | | 51 | Donor Blood Type B | 0.015 |
| | | | 52 | Donor Bilirubin | 0.012 |
| | | | 53 | Donor Weight | 0.000 |

**Table S4.** Feature used in ToPs sorted by normalized relevance score. (Features selected by CFSS are represented in gray.)

| No | Abbreviation | Meaning |
|---|---|---|
| 1 | ACE | Angiotensin-converting Enzyme |
| 2 | ACEI | Angiotensin-converting Enzyme Inhibitor |
| 3 | ARB | Angiotensin Receptor Blocker |
| 4 | AUC | Area Under the Curve |
| 5 | BMI | Body Mass Index |
| 6 | BNP | B-type Natriuretic Peptide |
| 7 | BUN | Blood Urea Nitrogen |
| 8 | COPD | chronic obstructive pulmonary disease |
| 9 | CVA | Cerebrovascular Accident |
| 10 | DRI | Donor Risk Index |
| 11 | ECMO | Extracorporeal Membrane Oxygenation |
| 12 | EF | Ejection Fraction |
| 13 | eGFR | Estimated Glomerular Filtration Rate |
| 14 | HEP | Hepatitis |
| 15 | HFSS | Heart Failure Survival Score |
| 16 | HLA | Human leukocyte antigen |
| 17 | IABP | Intra-Aortic Balloon Pump |
| 18 | IMPACT | Index for Mortality Prediction After Cardiac Transplantation |
| 19 | LVAD | Left Ventricular Assist Device |
| 20 | LVEF | Left Ventricular Ejection Fraction |
| 21 | MAGGIC | Meta-Analysis Global Group in Chronic Heart Failure |
| 22 | NYHA | New York Heart Association |
| 23 | pkVO2 | Peak Oxygen Consumption by Cardiopulmonary Exercise Testing |
| 24 | PRA | Panel Reactive Antibody |
| 25 | RSS | Recipient Stratification Score |
| 26 | RVAD | Right Ventricular Assist Device |
| 27 | VAD | Ventricular Assist Device |

**Table S5.** Abbreviations